# Design and Construction of VUES: the Vilnius University Echelle Spectrograph

Colby Jurgenson[†], Debra Fischer, Tyler McCracken, David Sawyer, Matt Giguere, Andrew Szymkowiak, Fernando Santoro[‡], and Gary Muller[‡]

[†]*Astronomy Department, Yale University, New Haven, CT 06511, USA, colby.jurgenson@yale.edu*
[‡]*ASTRO Electro-Mechanical Engineering, 1283 Carrizo St. NW, Los Lunas, NM 87031, USA*

In February of 2014 the Yale Exoplanet Laboratory was commissioned to design, build, and deliver a high resolution (R = 60,000) spectrograph for the 1.65-meter telescope at the Molėtai Astronomical Observatory. The observatory is operated by the Institute of Theoretical Physics and Astronomy at Vilnius University. The Vilnius University Echelle Spectrograph (VUES) is a white-pupil design that is fed via an octagonal fiber from the telescope and has an operational bandpass from 400 to 880 nm. VUES incorporates a novel modular optomechanical design that allows for quick assembly and alignment on commercial optical tables. This approach allowed the spectrograph to be assembled and commissioned at Yale using lab optical tables and then reassembled at the observatory on a different optical table with excellent repeatability. The assembly and alignment process for the spectrograph was reduced to a few days, allowing the spectrograph to be completely disassembled for shipment to Lithuania, and then installed at the observatory during a 10-day period in June of 2015.

## 1. Introduction

### 1.1. *The Molėtai Astronomical Observatory*

The Molėtai Astronomical Observatory is located about 70 km north of the city of Vilnius, Lithuania. It is operated by the Institute for Theoretical Physics and Astronomy at Vilnius University. It sits 200 meters above sea level just outside the town of Moletai, atop Kaldiniai Hill. There are three separate telescopes at the site with diameters of 0.51, 0.63, and 1.65 meters. The spectrograph discussed in this paper was designed and constructed for use on the f/12 1.65 meter Ritchey-Chretien telescope. The observatory building and the 1.65-m telescope are shown in the left and right panels (respectively) of Figure 1. In preparation for the installation of the spectrograph, the observatory re-aluminized the primary mirror, updated the telescope control system, and built a thermally controlled, light-tight room to house the instrument.

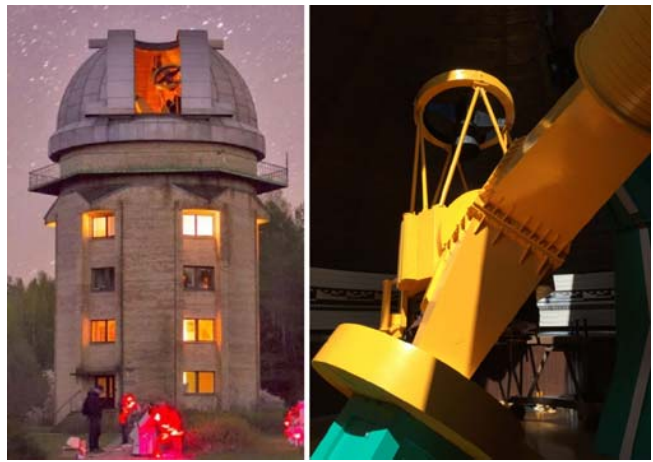

Figure 1 – *Left*: The MAO 1.65 meter telescope building. *Right*: The 1.65 meter, f/12 Ritchey-Chretien telescope that the spectrograph will interface to.

## 1.2. 1.65 Meter Telescope Specifications

The MAO 1.65-meter telescope is a f/12 Ritchey-Chretien design. The optics were fabricated in 1976 at the Lytkarino Optical Glass Factory in the Moscow region of Russia. The telescope has been in operation since 1991. Table 1 lists some of its specifications.

Table 1. The MAO 1.65 meter telescope specifications.

| Parameter | Specification |
|---|---|
| Telescope Architecture | Ritchey-Chretien |
| Primary Diameter | 165 cm |
| Primary Focal Length | 492 cm |
| Secondary Diameter | 45 cm |
| Focal Ratio | 1:12 |
| Equivalent Focal Length | 1968 cm |
| Field of View | 35 arcminutes |

## 1.3. Spectrograph Design Requirements

In the fall of 2013 the Yale Exoplanet Laboratory responded to a call for proposals to design, build, and deliver a high-resolution spectrograph for MAO. The contract was formalized in February of 2014, and the instrument was delivered in June of 2015. Table 2 lists the high-level requirements for the instrument. VUES is a general-purpose spectrograph intended to be used for stellar spectroscopy.

Table 2. High-level instrument requirements for the MAO spectrograph instrument package.

| Parameter | Specification |
|---|---|
| Wavelength Range | 400 to 880 nm |
| Spectral Resolution Modes | 30,000/45,000/60,000 |
| Spectral Sampling | 4 pixels for the highest resolution mode |
| Instrumental Throughput | 25% at 543 nm |
| On-sky Fiber Aperture | 2.5 arceseconds |
| Spectrograph Enclosure Temperature Stability | ± 1C over 12 hours |
| Front-End Module Operational Temperature Range | -30C ≤ T ≤ +30C |
| Spectrograph Detector | 4k x 4k x 15 µm pixel pitch |
| Instrument Control Software | GUI based observer interface |
| Data Reduction Software | Quick-look assessment & full data extraction pipeline |

## 1.4. Science Case

VUES is to be used primarily for studies of physical parameters and abundance analysis of evolved stars belonging to red giant, asymptotic and post-asymptotic giant branch stars (RGB, AGB, and post-AGB). The instrument will be dedicated to long-term research programs to investigate thermally pulsating AGB and post-AGB stars complemented by simultaneous broadband BVR photometry taken at the observatory. Long term monitoring of nearby K-M dwarf spectroscopic binaries, originally undertaken from 2000 to 2015 with a Coravel-type spectrometer at the observatory, will be continued as well.

## 1.5. Instrument Architecture

The VUES instrument architecture is shown schematically in Figure 2. There are two distinct modules that house multiple optical subsystems; the front-end module (FEM) and back-end module (BEM). The FEM interfaces directly to the telescope Cassegrain port. The f/12 telescope is focused onto a mirror with a 2.5 arcsecond diameter hole (matched to the typical seeing), which allows light from the science target to pass through. The remainder of light is reflected by the 5 arcminute diameter mirror and is reimaged onto a camera at f/4. This camera has two functions: 1) initial target acquisition and centering, and 2) positional error feedback to the telescope mount at a rate of 1 Hz. The software allows for guiding to either take place on the spill-over light from the on-axis target star, or on an off-axis field star within the 5 arcminute field of view.

The 2.5 arcsecond target field is passed into the fiber feed optics. These consist of a pair of custom designed achromats that convert the native f/12 beam from the telescope to an f/5. There is a 2.5 arcsecond diameter octagonal fiber located at the f/5 focus. It collects the light and then propagates it to the spectrograph input. The octagonal fiber was chosen for its superior scrambling properties (Spronck et al. 2012). The calibration unit is located near the BEM and light from the quartz or thorium argon lamps is brought up to the FEM with a circular fiber and injected into the octagonal science fiber with optics that provide the correct f/5 focal ratio.

After the octagonal fiber, the f/5 beam is reimaged onto a slit mask at f/10 before being fed into the spectrograph. The remotely controllable slit mask has three positions that provide resolutions of 30,000, 45,000, or 60,000. The R=30,000 slit mask aperture is matched to the image size from the fiber and the higher resolution modes vignette the sides of the image and incur some light loss. The slit mask also includes a pinhole that was used in aligning the instrument. The spectrograph is a white-pupil configuration with a custom f/6 camera that is optimized to operate between 400 and 880 nm. Section 2 will discuss the optical design, Section 3 the mechanical design, Section 4 the calibration injection, Section 5 instrument control and data reduction, and in Section 6 laboratory results will be presented.

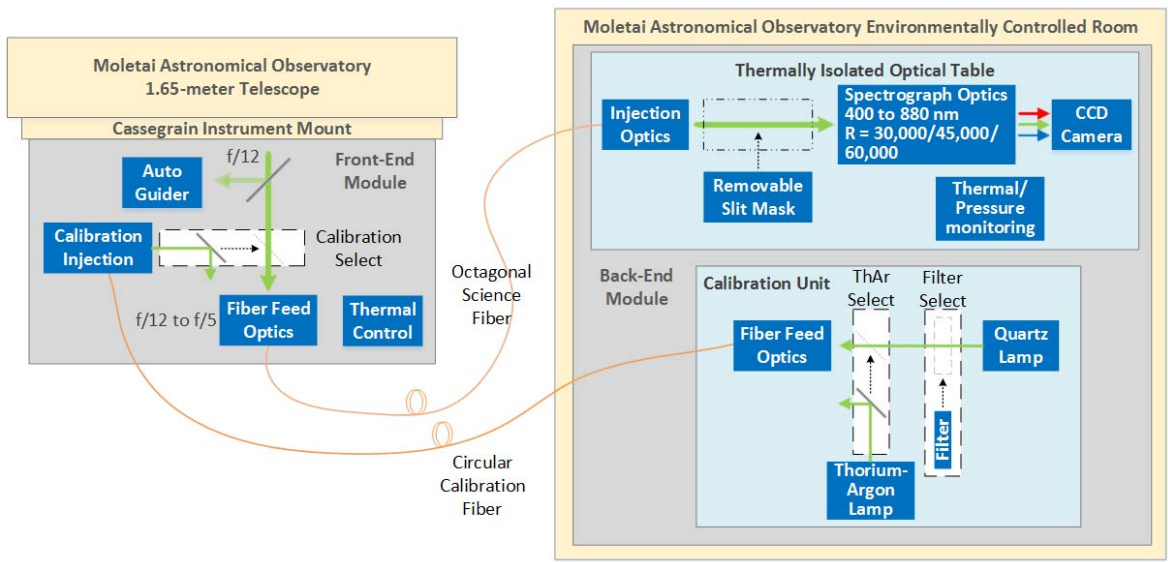

Figure 2 – Instrument architecture of the Yale designed and built high-resolution spectrograph for the MAO 1.65 meter telescope.

## 2. Optical Design

As depicted in Figure 2, and described in Section 1.4, the MAO instrument package is comprised of two modules; The Front-End Module (FEM), and the Back-End Module (BEM). Section 2.1 describes the details of the optical design of the FEM, and Section 2.2 describes the BEM design.

### 2.1. *Front-End Module*

The FEM can be broken out into three distinct folded light paths or arms; these are highlighted in the three panels of Figure 3. In the left panel, the f/12 beam from the telescope is reflected toward the autoguider arm; the focal plane mirror contains the central 2.5 arcsecond hole for transmitting light to the octagonal fiber. The light is folded again for alignment of the fiber to the hole in the focal plane mirror. The center panel of Figure 3 is a plan view of the autoguider and calibration injection arms.

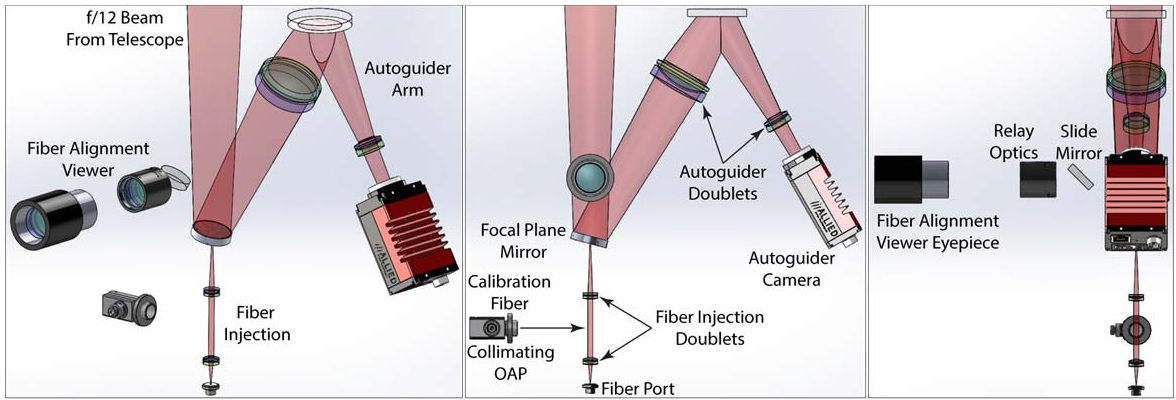

Figure 3 – *Left*: The three folded light paths of the FEM; autoguider, fiber injection, and fiber alignment. *Center*: Plan view of the autoguider and fiber injection arms. *Right*: Plan view of the fiber alignment arm.

### 2.1.1. Autoguider

Light from the telescope is first intercepted by the focal plane mirror (see center panel of Figure 3). This mirror is made from highly polished aluminum with a 240 μm laser drilled hole in the center of it. The hole is drilled at 15° relative to the surface normal so that the central 2.5 arcseconds passes into the fiber injection arm, and the 5 arcminute field is reflected into the autoguider arm. Figure 4 is an image of the focal plane mirror taken with the autoguider camera. The central hole can be seen as the dark spot in the center. The bottom of the figure is a plan view of the mirror showing the laser drilled hole, and conical shaped cut-away from the interior that allows the 2.5 arcsecond target field to pass unobstructed to the fiber injection. This mirror was fabricated by Lenox Laser.

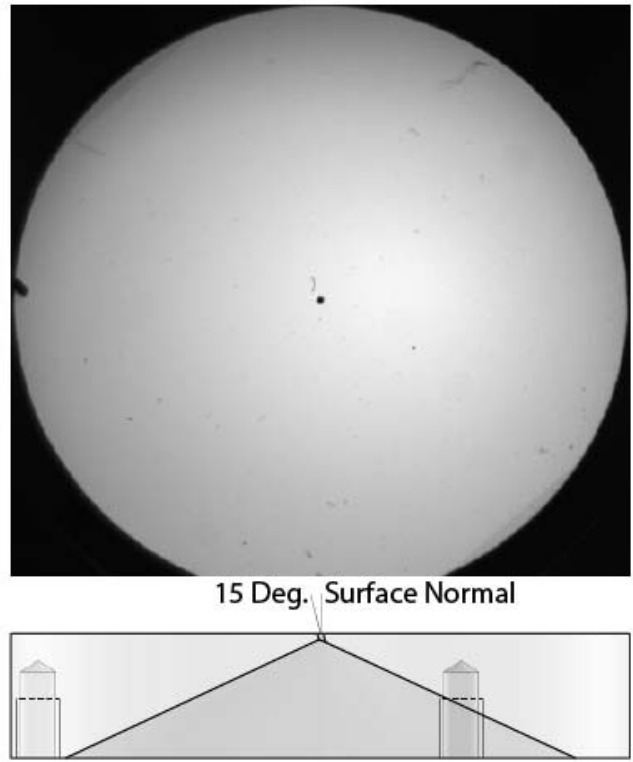

Figure 4 – *Top*: An image of the focal plane mirror taken with the autoguider camera during focusing. *Bottom*: Plan view drawing of the mirror showing the laser drilled hole and conical shaped cut-away of the mirror interior that allows the target field to pass to the fiber injection.

The autoguider detector is a Prosilica GT 2750 with a 6-megapixel type-1 CCD that allows for up to 20 frame readouts per second. An additional feature of this camera is that it can operate over a temperature range of -20C to +60C; this was a requirement given the ambient temperature conditions at the MAO site. As noted in Table 2, the ability to operate over this temperature range is a requirement for all optomechanical components in the FEM. The autoguider camera is used for target acquisition, and positional offset computation for telescope mount corrections at rates up to 1 Hz. Guiding can be carried out by either centroiding on the spill over light from the on-axis target star, or on an off-axis field star. Two custom achromatic doublets were fabricated by Rocky Mountain Instruments (RMI) to convert the f/12 beam from the telescope to f/4 on the detector. Custom anti-reflection (AR) coatings were applied to all transmissive optics in the FEM with an average reflectance of <1% over the full 400 to 880 nm. Figure 5 is an optical ray trace of the autoguider arm with the doublet lens materials and the focal plane mirror and detector plate-scales are indicated.

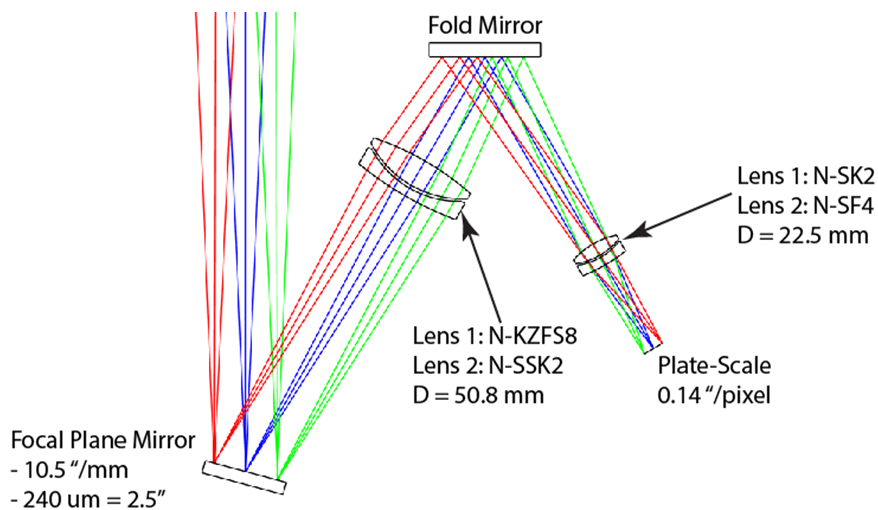

Figure 5 – Optical ray trace highlighting the autoguider arm design. Lens materials as well as the focal plane mirror and detector plate-scales are called out.

### 2.1.2.  *Fiber Injection*

The fiber injection arm takes the light that passes through the focal plane mirror and injects it into the fiber that feeds the spectrograph. These optics are two custom achromatic doublets that convert the native f/12 from the telescope to f/5 at the fiber. Figure 6 is an optical ray trace of the fiber injection optics. Light is propagated to the spectrograph with a 100-μm octagonal Polymicro broad-spectrum optical fiber (FBPI) with AR-coated faces. There is a collimated space between the doublets that allows for calibration light to be injected into the fiber.

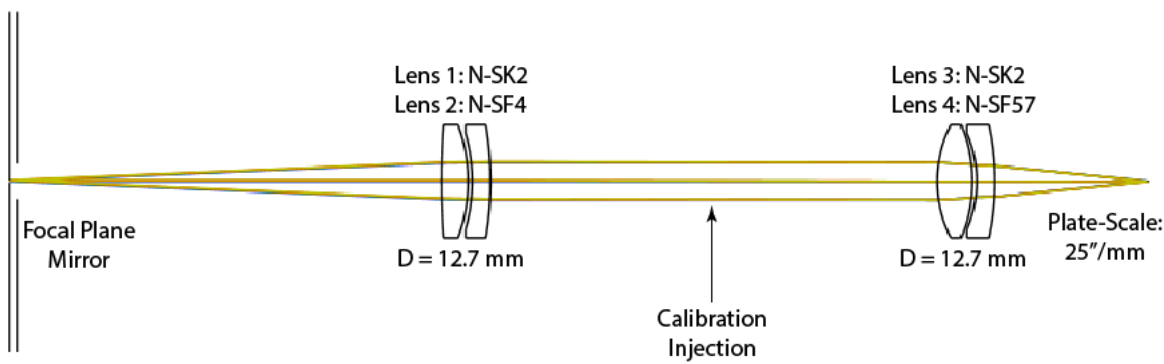

Figure 6 – Optical ray trace of the fiber injection doublets. Light from the calibration source fiber is inserted via flip mirror in the collimated space between the doublets.

This is done with an off-the-shelf Thorlabs fiber OAP collimator, and a flip-in mirror (not shown). The fiber injection doublets were also manufactured by RMI with the same AR coatings. Best focus for both the autoguider and fiber injection arms of the FEM is achieved by moving the telescope secondary mirror. An Invar-based mechanical structure was designed and fabricated to maintain the depth of focus for both arms over the entire temperature range requirement (Section 3).

### 2.1.3. *Fiber Alignment*

The right panel in Figure 3 is a side view of the FEM layout highlighting the fiber viewing and alignment system, and built with off-the-shelf components. Light from a green LED source is back propagated from the spectrograph through the octagonal fiber to the FEM. The fiber viewing and relay lens pick-off mirror are slid into place above the focal plane mirror, forming an image of the relative positions of the central hole and fiber as viewed through an eyepiece. The eyepiece itself is mounted into an Orion telescope focuser and bolted to the side of the FEM enclosure. The relay optics have a 1 to 2 magnification ratio, and the eyepiece is a 30x Plossl that magnifies the 100 μm fiber to 1.5 mm. To align the fiber to the central hole in the focal plane mirror, the observer simply adjusts the fiber port until the glow from the reimaged fiber head is centered in the hole.

## 2.2. *Back-End Module*

As defined in Section 1.4, the BEM consists of the spectrograph and the calibration unit. It interfaces to the FEM via two fibers; the octagonal science fiber brings light from the telescope to the spectrograph and a circular fiber transfers calibration light from the BEM to the FEM. Both of these units reside in an environmentally controlled room beneath the observing platform. The calibration unit will be discussed separately in Section 4.

### 2.2.1. *Spectrograph*

A white pupil configuration (Baranne 1972) was selected for the spectrograph architecture, as illustrated in Figure 7. Light exits the fiber at f/5, and is then reimaged at f/10 at the slit mask. The reimaging optics consist of two off axis parabolas (OAPs) from Space Optics Research Labs with protected silver coatings to increase reflectance in the blue end of the spectrum. Following the slit mask the beam expands until it is 80 mm in diameter at the main collimator OAP, where it is reflected to an R4 echelle grating.

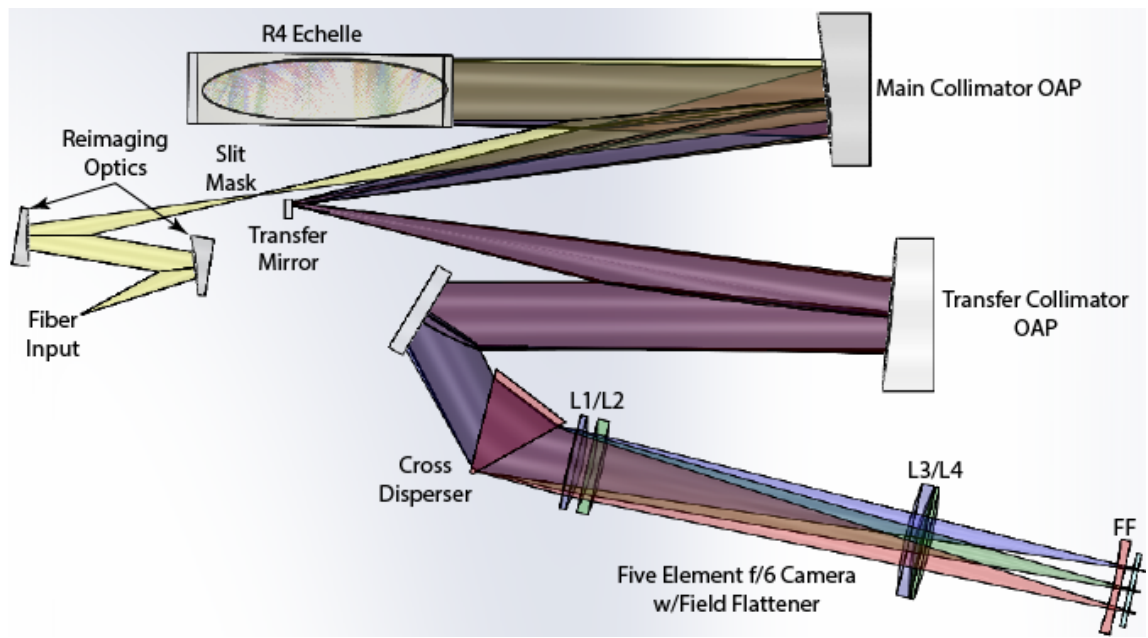

Figure 7 – The MAO white pupil spectrograph layout.

The grating was fabricated by Newport Richardson Grating Laboratory, and has a protected silver coating with 31.6 grooves per mm. An off-plane tilt of 0.75° is applied to the grating in order to displace the focus of the collimating OAP on second pass. After reflecting off the echelle, the collimating OAP focuses the now dispersed beam. The transfer mirror intercepts the beam 40 mm before focus and directs it to the transfer collimator OAP where the beam is recollimated before being cross-dispersed by a prism. Both the main and transfer collimators are identical in terms of parent radius and off-axis displacement. They are also standard catalog products offered by Space Optics Research Labs and have protected silver coatings.

The white pupil following the transfer collimator was placed on the first camera lens. Except for the cylindrical field flattener, the camera is an all-spherical, modified Petzval design with a system focal ratio of f/6. All lens materials were selected from the Schott preferred catalog to include glasses with good transmission in the bluer end of the spectrum. The one exception is the first lens, where the best optical properties were from calcium fluoride. Every surface is AR coated with the same dielectric coating as the FEM optics that achieves an average reflectance of <1% from 400 to 880 nm. Table 3 lists the specifications for the cross disperser and camera optics.

Table 3 – Optical specifications of the cross disperser and camera optics.

| Element | Material | Surface 1 Radius (mm) | Surface 2 Radius (mm) | Thickness (mm) | Diameter (mm) |
|---|---|---|---|---|---|
| Cross Disperser | N-SF57 | NA | NA | 60 | 130 (length/height) |
| L1 | Calcium Fluoride | 809.71 | 238.08 | 19.07 | 127 |
| L2 | SF1 | 229.38 | 452.47 | 12.70 | 127 |
| L3 | N-LAK10 | 476.88 | 269.05 | 12.83 | 152.4 |
| L4 | SF2 | 329.06 | 540.30 | 19.14 | 152.4 |
| FF | SF6 | 479.43 (x-radius) | Infinite | 9.64 | 127 x 127 (Square) |

The near diffraction-limited performance of the camera is shown in Figure 8. Each curve in the figure is the diffraction ensquared energy for a different spectral order. It was computed by taking the average of five wavelengths that span the given order. A total of eleven orders are plotted across the entire wavelength range from 400 to 880 nm.

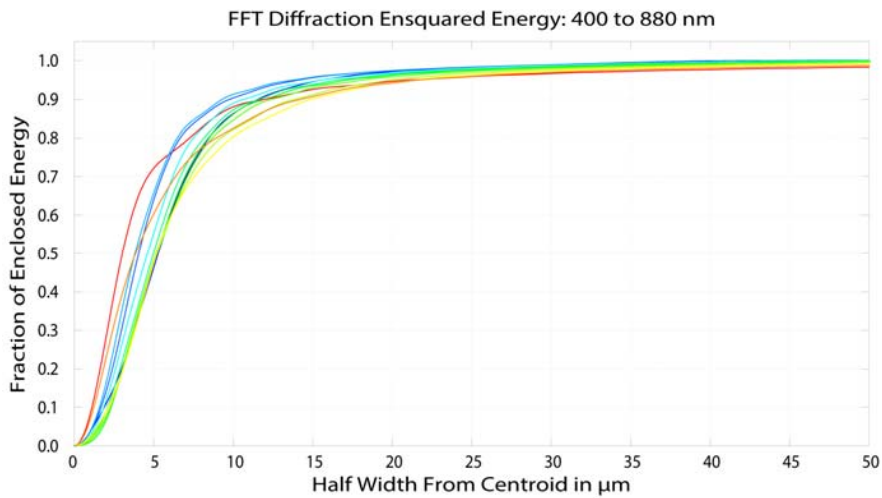

Figure 8 – The diffraction ensquared energy plotted for eleven orders across the entire wavelength range.

## 3. Mechanical Design

The Yale Exoplanet Laboratory contracted with ASTRO Electro-Mechanical Engineering of Los Lunas, New Mexico to perform the mechanical design and analysis of the FEM and BEM. Figure 9 is a three-dimensional rendering of the spectrograph optomechanical design. Section 3.1 will give an overview of the FEM mechanical design, and Section 3.2 will describe that for the BEM.

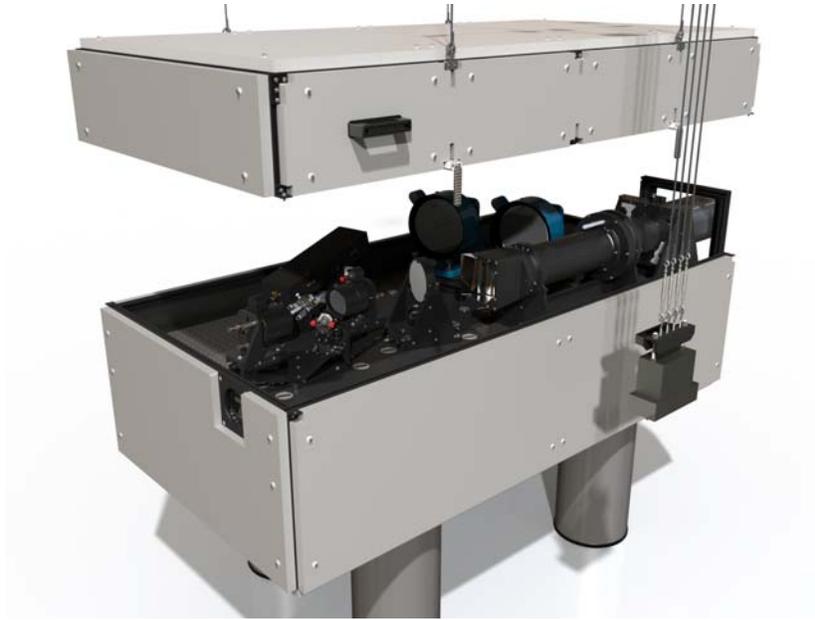

Figure 9 – Three dimensional rendering of the spectrograph optomechanical design.

### 3.1. *FEM*

The FEM houses the optomechanics that collect the light from the telescope and inject it into the fiber for transfer to the spectrograph. It also provides the optomechanics to inject light from the calibration unit into the science fiber. The left panel of Figure 10 is a three-dimensional rendering of the FEM structural assembly. There are three main parts to it: 1) upper structural support (telescope interface), 2) the lower structural support, and 3) the fiber port cover. The right panel of Figure 10 is an image of the as built FEM installed on the back of the MAO 1.65 meter telescope. The entire FEM has an approximate mass of 25 kg, and occupies a 553 x 430 x 430 mm volume.

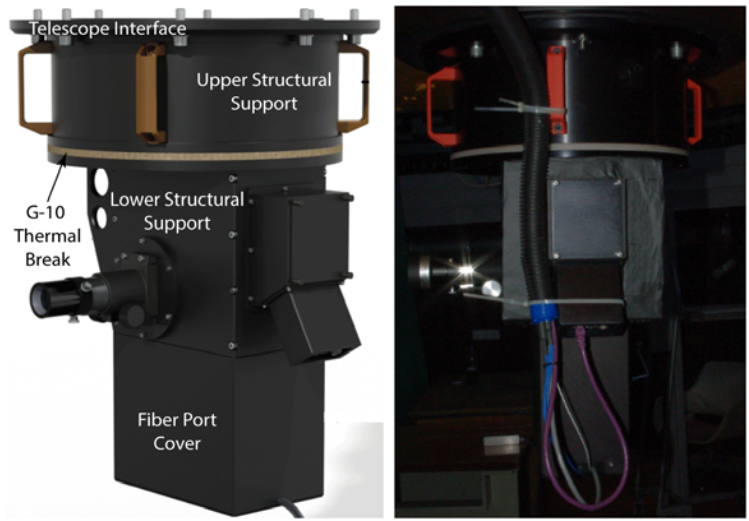

Figure 10 – *Left*: A three dimensional rendering of the FEM structural support assembly. *Right*: The as built assembly mounted on the MAO telescope.

### 3.1.1. *Structural Support Assembly*

The structural support assembly mounts to the telescope and supports all the optomechanics that make up the FEM. Figure 11 highlights the various design features of the FEM. It is an all-aluminum construction and provides the interface to the telescope, and the stiff spacer tube optomechanics. The telescope mounting flange is welded to the upper structural support. The lower structural support is a bolted structure that locates the optomechanics, and provides access to them. Due to the humid environment at the observatory site, resistive heating elements were added to the inner walls of the lower structural support. A G-10 fiberglass plate between the upper and lower structural assemblies is used to break the thermal path. Removable insulation panels were constructed that can be attached to both the lower support and fiber port cover.

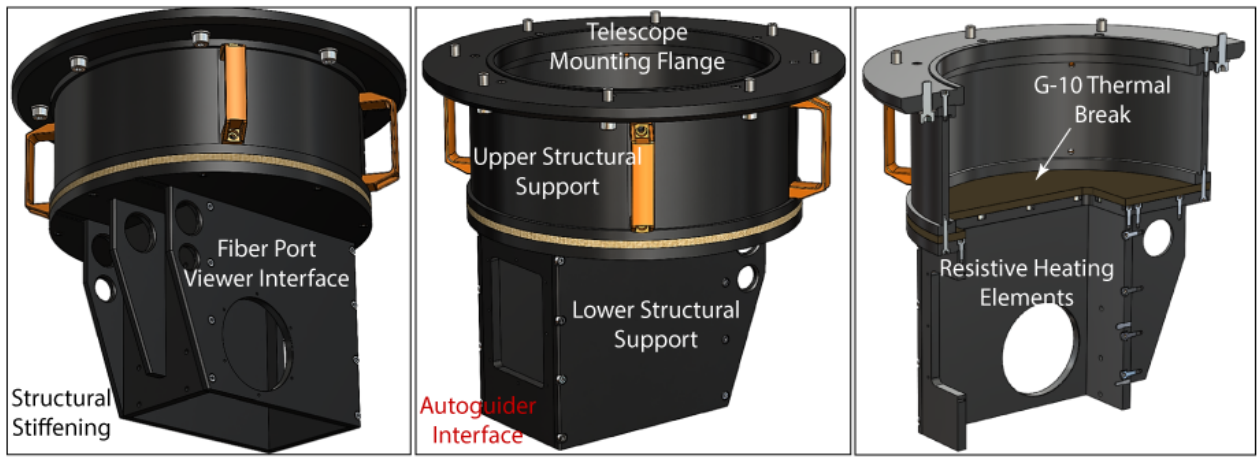

Figure 11 – Left: Isometric view of the structural support assembly highlighting the structural stiffening and fiber port viewer interface. Center: Isometric view highlighting the different sections and autoguider interface. Right: Cut-away isometric view highlighting the G-10 interface between the upper and lower structural supports.

### 3.1.2. *Main Optomechanical Assembly*

Figure 12 shows renderings of the main optomechanical assembly. This assembly houses the focal plane mirror, autoguider, and fiber viewing elements (not shown). The main Invar 36 optics mounting frame is thermally isolated from the lower structural support by a G-10 fiber glass plate. At the aluminum/G-10/Invar interface, bullet nose and diamond pins allow differential expansion and contraction due to the different material coefficients of thermal expansion.

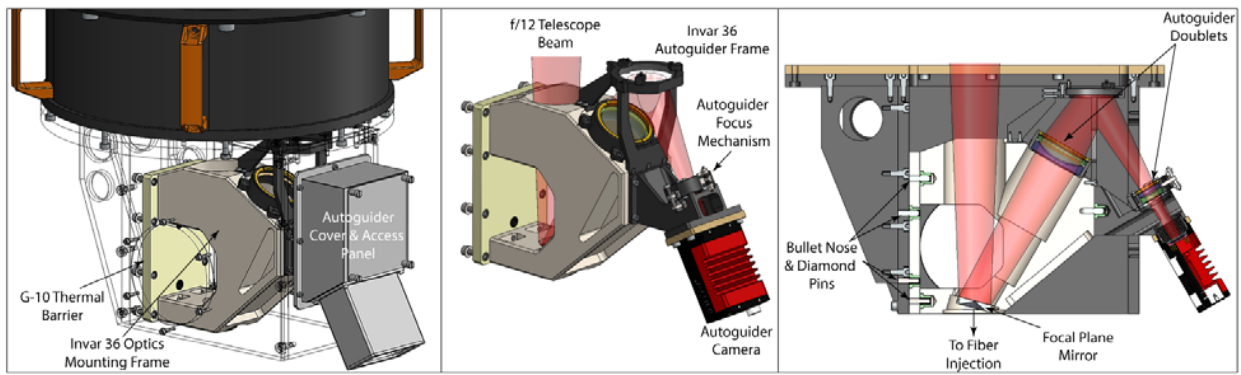

Figure 12 – Left: The main optomechanical assembly highlighted with the lower structural transparent. Center: The main optomechanical assembly showing the autoguider Invar 36 frame. Right: Autoguider optical path in plan view.

The second pair of autoguider doublets are mounted in a housing that employs a flexure-based focus assembly. As was demonstrated in Figure 4, the autoguider camera can be focused on the pinhole mirror. This mechanism is then locked down, and autoguider focus is maintained by adjusting the telescope secondary mirror focus mechanism. From finite element analysis (FEA), it was shown that the path length from the focal plane mirror to the autoguider detector will not change by more than 27 µm. This is within the system depth of focus of 38 µm.

### 3.1.3. Fiber Port Assembly

The fiber port assembly bolts onto the main optomechanical assembly immediately following the focal plane mirror. This arrangement is shown in the panels of Figure 13. The assembly has three main subcomponents: 1) fiber relay, 2) calibration unit injection, and 3) the fiber port. As discussed in Section 2.1.2, L1/L2 is a custom doublet. Its assembly is fixed relative to the focal plane mirror. The L3/L4 lenses are mounted in a flexure mechanism to focus the light being injected into the fiber during alignment. Once it is aligned and focused, the mechanism is locked down, and focus is accomplished using the telescope secondary mechanism. The assembly material is Invar 36, and FEA has shown that the optical path between the focal plane mirror and fiber will not change by more than 10 µm over the -30C to +30C temperature swings measured at the site. The depth of focus of L3/L4 is 60 µm.

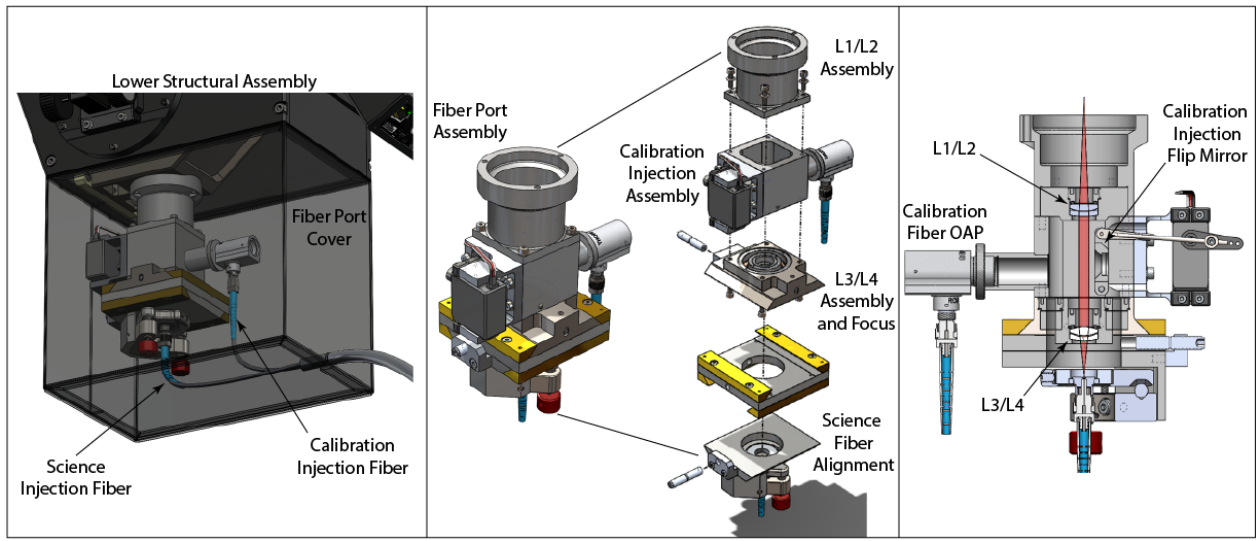

Figure 13 – *Left*: Rendering with transparent fiber port cover showing the fiber port assembly attached to the main optomechanical assembly. *Center*: Fiber port assembly broken out into its subcomponents. *Right*: Cut-away plan view highlighting the optical components in the assembly.

The fiber port itself is an off-the-shelf Thorlabs tip/tilt mount attached to a pair of dovetail slides that provide x/y-adjustment. The combination of the tip/tilt mount and the dovetail slides allow for all tilt and shear alignment errors between the fiber axis and the optical axis to be removed. A micro-linear actuator allows for a mirror to be flipped into the beam path between L2 and L3 to inject calibration light into the fiber. An off-the-shelf collimating OAP fiber port from Thorlabs takes light from the calibration fiber and directs it toward the flip mirror. The mirror is seated by an adjustable hard stop to ensure repeatable positioning.

### 3.2. *Spectrograph*

Figure 14 is a picture taken of the spectrograph optical bench and its components in the climate controlled room at the MAO. The lid of the enclosure has been removed from its base, exposing the optomechanics for the picture. The different components can be divided up into five groupings: 1) alignment templates, 2) fore-optics group, 3) mid-optics group, 4) aft-optics group, and 5) the enclosure. Each of these groupings will be discussed in the following sections.

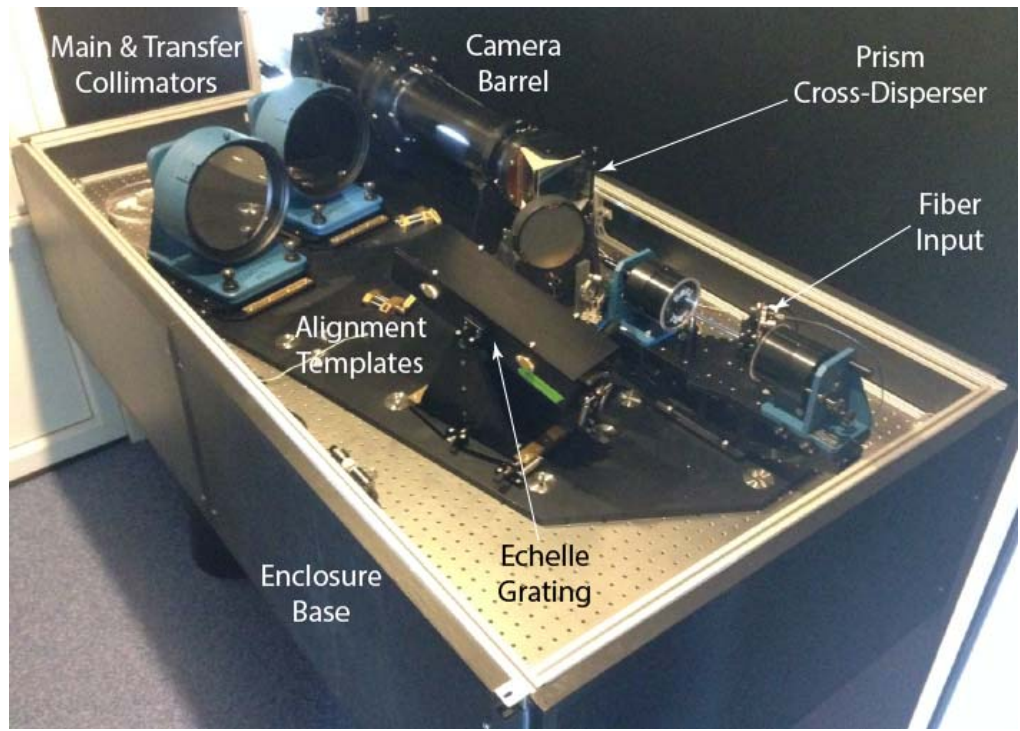

Figure 14 – The spectrograph optical bench and components installed at the MAO in its climate controlled room.

### 3.2.1. *Alignment Templates*

Optomechanical components are divided into three distinct groups (see Sections 3.2.2/3.2.3/3.2.4), each group with its own template. Templates mount to the table, the optical mounts interface with the templates, and the templates interface with each other via two precision contact points (alignment pins). The templates are then fastened together using the template fasteners. This approach has been successfully applied for an interferometric beam combining instrument (Jurgenson et al. 2013).

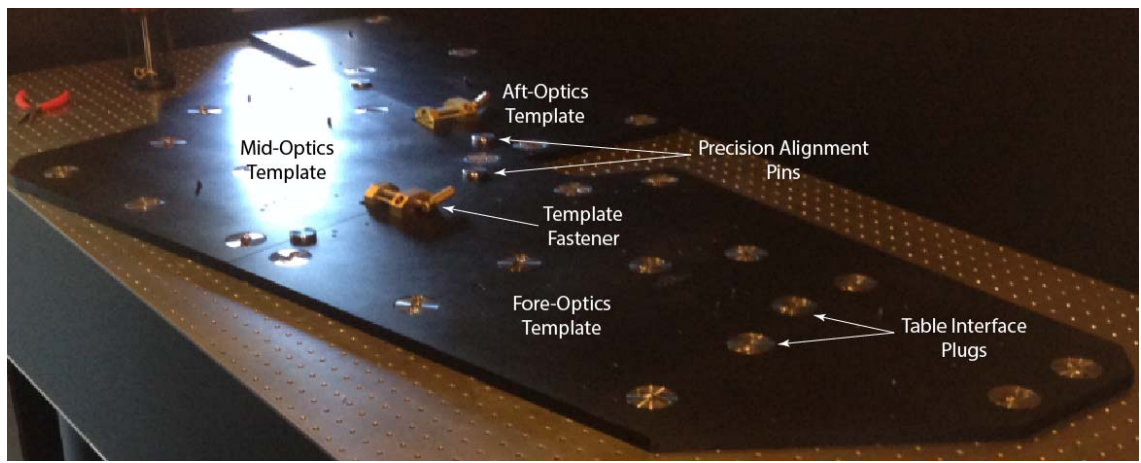

Figure 15 – The alignment templates installed on the VUES optical bench.

This approach is advantageous from an alignment perspective because it positions the mounts relative to each other to machine tolerances. Therefore the instrument can be quickly assembled into a near-optimal alignment state. It also increases the long term optomechanical stability because it can help to reduce the number of degrees of freedom for alignment designed into each mount. For high-resolution spectrographs that do not require extreme precision, this solution offers stability without a signficant expense in terms of cost or technology. Figure 16 is a plan view of the spectrograph optomechanical design showing the three templates and the optical groupings that reside on each.

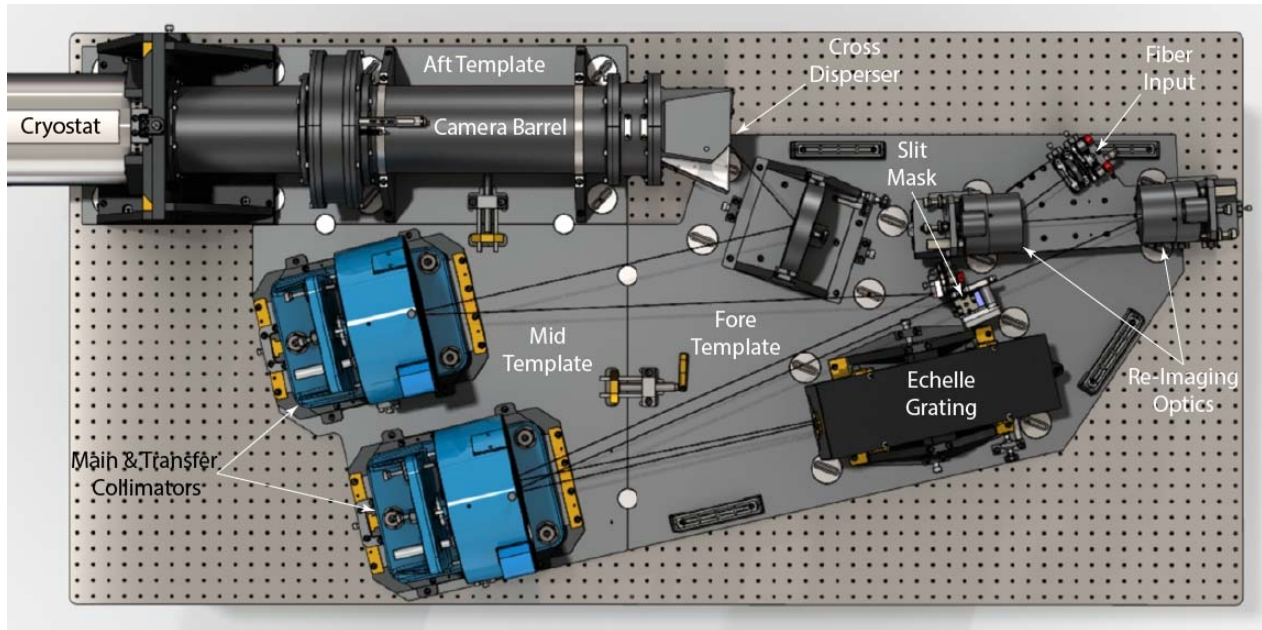

Figure 16 – Overview of the alignment templates and optics groupings that reside on each. The black line traces the beam path.

### 3.2.2. *Fore-Optics Group*

The principal components of the fore-optics group are the fiber input, reimaging optics, slit mask assembly, and the echelle grating. Figure 17 shows renderings of the fiber port mount and the slit mask assembly mount. These mounts interface to the template using precisely located bullet nose pins and bushings. Light exits the fiber and is then reimaged onto the slit via OAP-1 and -2.

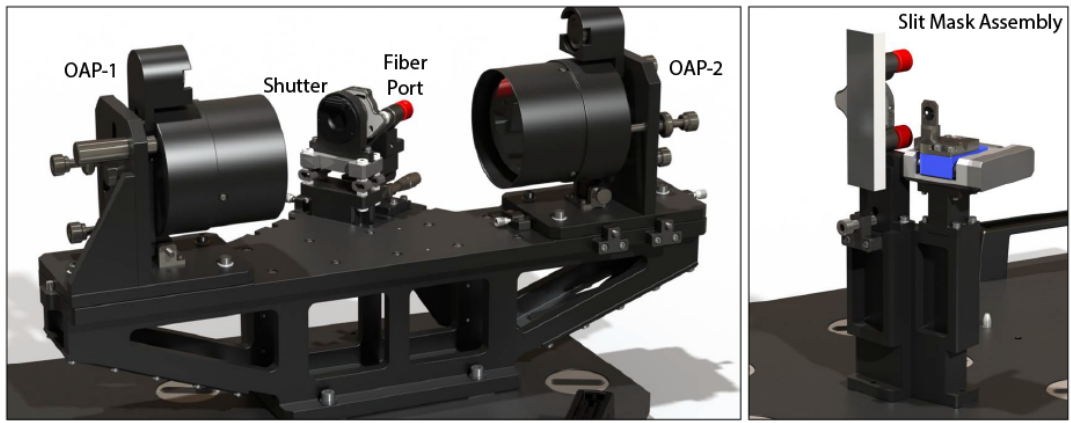

Figure 17 – *Left*: The fiber port, shutter, and reimaging OAPs all share a common mount interface to the template. *Right*: The slit mask assembly includes the transfer mirror.

The location of the slit is what we have termed the "principle point" of the spectrograph, meaning that it is fixed with no adjustments, and therefore everything is aligned to it. The slit mask has four positions corresponding to the three different resolutions, and a pinhole that is used during the alignment process. OAP-1 is fixed in place to machine tolerances, and the fiber port has tip/tilt and translation to align its output to the fixed OAP axis. OAP-2 can then be translated and to place the focus at the slit. Alignment results will be discussed in greater detail in Section 6. Figure 18 shows two detailed renderings of the echelle grating mount. The echelle mount is located on the template using a precisely located pin and bushing.

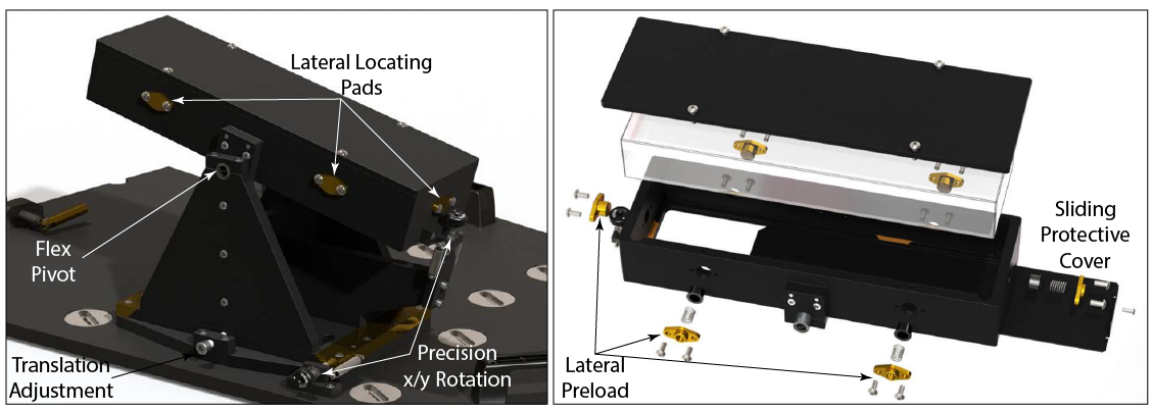

Figure 18 – *Left*: The echelle grating mount on fixed to the template. *Right*: Features of the echelle enclosure.

### 3.2.3. *Mid-Optics Group*

The mid-optics group consists entirely of the main and transfer collimating OAPs and their x/z translation stages. These are shown in the left and right panels of Figure 19. The OAPs and the transfer collimators are standard catalog optics offered by the Space Optics Research Labs. The optics and mounts arrive separately, so the optics must first be aligned within the mounts and then the mounts are aligned to the spectrograph optical axis. To accomplish the latter, the mounts are fastened to x/z translation stages. These stages are two dovetail slides stacked on top of each other with push-pull screws. The z-translation allows the OAP to be focused, and the x-translation is to properly position the OAP off-axis distance with the spectrograph optical axis.

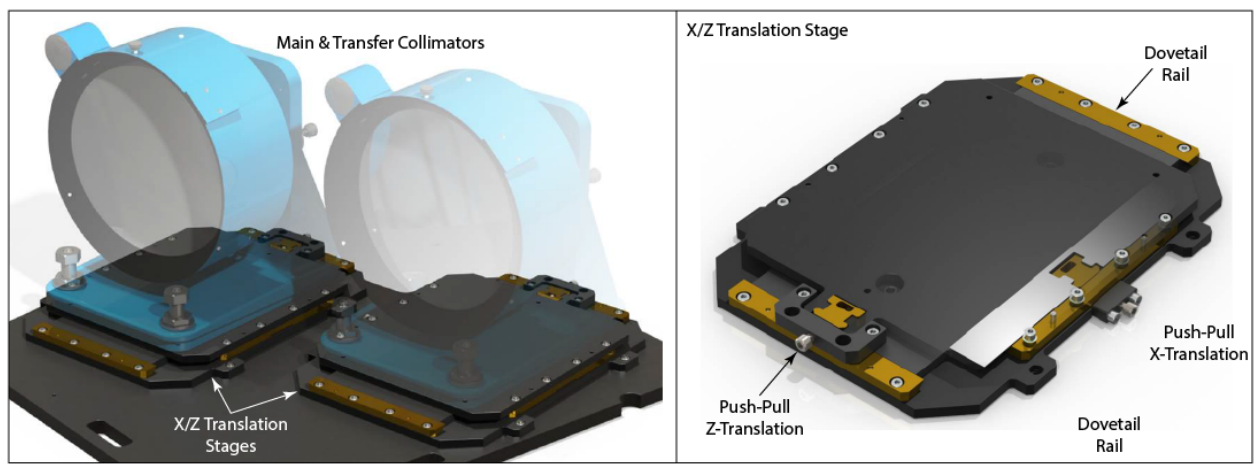

Figure 19 – Left: The main and transfer collimators (slightly transparent) mounted on their x/z translation stages which are mounted on the alignment template. Right: An x/z translation stage consists of two dovetail slides with push-pull screws.

### 3.2.4. *Aft-Optics Group*

The aft-optics includes four sub-groupings which are the cross-disperser, doublet L1/L2, doublet L3/L4, and the field flattener. Each of these sub-groupings are placed into individual mounts that interface to the other sub-groupings via a barrel structure. The barrel structure fixes and maintains the concentricity from L1 to the field flattener, and builds in the angular relationship between the cross-disperser and the barrel optics. There are two brackets that interface the barrel structure to the alignment template. One located near L1/L2, and the other at L3/L4. This is illustrated in Figure 20. Alignment template pin locations are called out for interfacing of the camera group template to the templates of the cryostat and main optics. The red line in the figure represents the optical axis of the spectrograph.

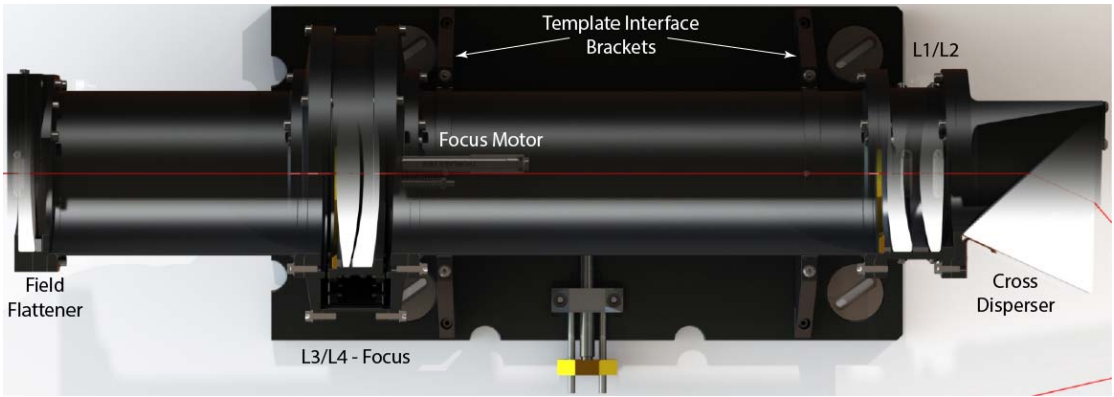

Figure 20 – The aft-optics group consists of the cross-disperser, L1/L2, L3/L4 doubles, and cylindrical field flattener. The L3/L4 doublet has a flexure mount to allow for focus. The red line represents the spectrograph optical axis.

The three panels in Figure 21 are renderings of the sub-grouping mounts for L3/L4, L1/L2, and the cross-disperser. Common to all mounts are bullet-nose pins at the interface to the barrels. This insures alignment between the different sub-groups. The individual doublet mounts maintain the proper spacing between the optics, as well as their internal concentricity. The L3/L4 mount has an additional degree of complexity in that it has a flexure stage for focus. L3/L4 are housed in an internal mount that then interfaces to a flexure structure that allows it to translate relative to the barrel structure. The flexure structure provides ±3 mm of travel while maintaining the concentricity of the mount to the barrel, and therefore the other sub-groupings of optics.

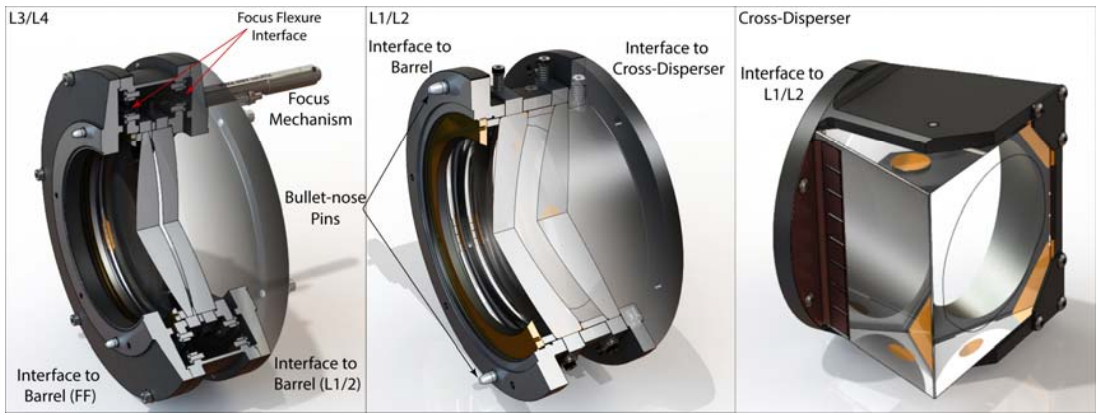

Figure 21 – Left: The L3/L4 mount. A focus flexure mechanism is built in to the mount and provides ±3 mm of travel while maintaining concentricity to the barrel structure. *Center*: The L1/L2 mount. Bullet-nose pins at all of the barrel interfaces insure alignment of the sub-grouping of optics relative to one another. *Right*: The cross-disperser mount insures the rotational position of the prism relative to the optical axis of the system.

### 3.2.5. *Cryostat Interface*

The cryostat is technically mounted on the aft-template, but it has its own degrees of freedom for adjustment and independent of the camera barrel. There are individual degrees for freedom for its alignment relative to the camera barrel in both translation, and rotation about the optical axis. Linear travel of ±5 mm is allowed for in both the x and y directions (z being the along the optical axis). The rotational axis has two stages of adjustment, a coarse stage with positions in 5° increments, and a fine stage that allows for 0.1° incremental adjustments over ±2.5°. This allows for fine coverage over the full 5° coarse rotational positions.

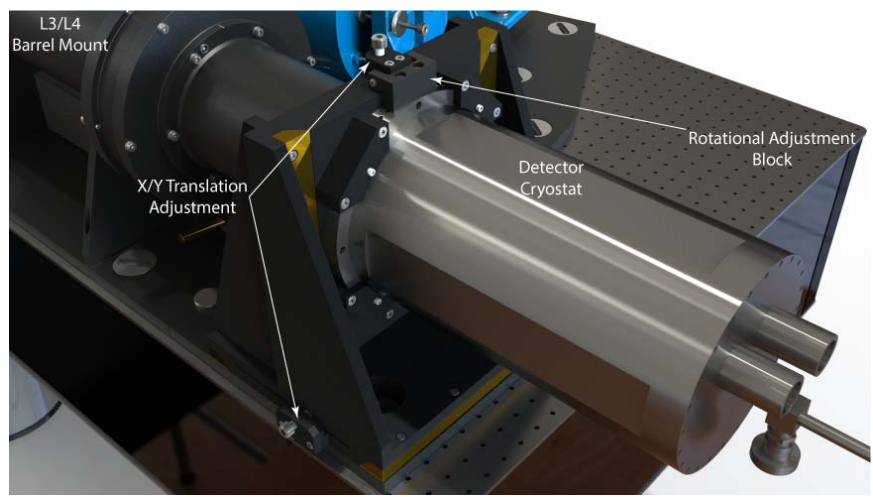

Figure 22 – The detector cryostat is mounted on the aft-template, but has translation and rotation degrees of freedom, independent from the camera barrel.

### 3.2.6. *Enclosure*

A custom enclosure was designed for the spectrograph. It was constructed out of 80-20 aluminum framing with Ultra-Board panels. The Ultra-Board panels have a rigid polystyrene foam core faced on both sides by a smooth, moisture resistant sheet of solid polystyrene. Each panel is one inch thick. The optical table is completely enclosed by the panels, and has an upper lid that is removable via a pulley system to provide global access to the spectrograph bench. It also provides a near light-tight environment as was demonstrated when spectra were taken with the lights in the room on and stray light was virtually undetectable.

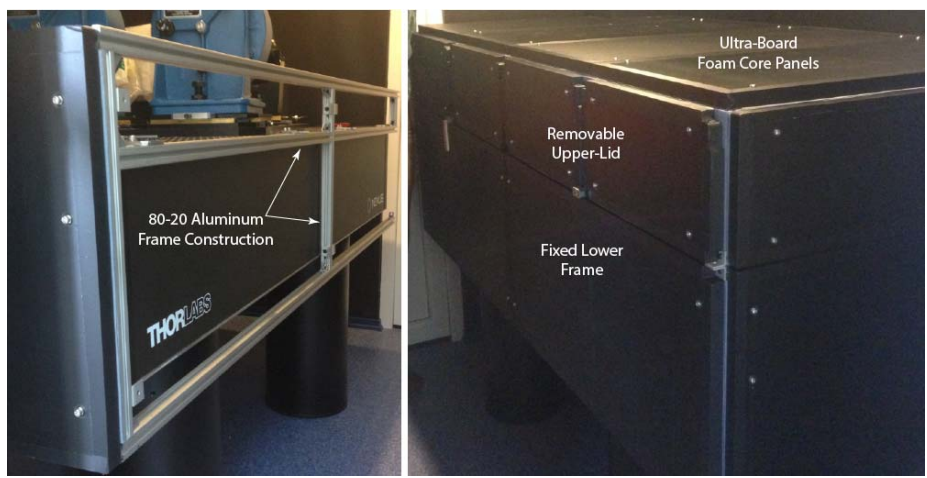

Figure 23 – The custom built enclosure for the spectrograph.

## 4. Calibration Injection

Calibration sources are required for removal of instrumental signatures in the data, and wavelength mapping of the spectrum. Wavelength dependent instrumental signatures are removed using a filtered quartz continuum source "flat-field" to measure and divide out any pixel-to-pixel variations in the CCD images along the orders of the echelle spectrum. A thorium argon (ThAr) lamp produces a known emission line spectrum for defining the wavelength mapping. These calibration sources are injected into the spectrograph through the science fiber to follow the same path as the science light.

### 4.1. Optomechanical Architecture

The calibration unit is part of the Back-End Module (BEM) and is located in the temperature-controlled room for lamp stability. The calibration unit contains ThAr and quartz calibration lamps, the optics for injecting the calibration sources into a fiber, and filters to flatten the quartz continuum. Figure 24 shows the optical layout of the calibration unit. A computer controlled flip mirror is used to select between the ThAr or quartz sources and a computer controlled filter mechanism inserts or removes a BG38 filter.

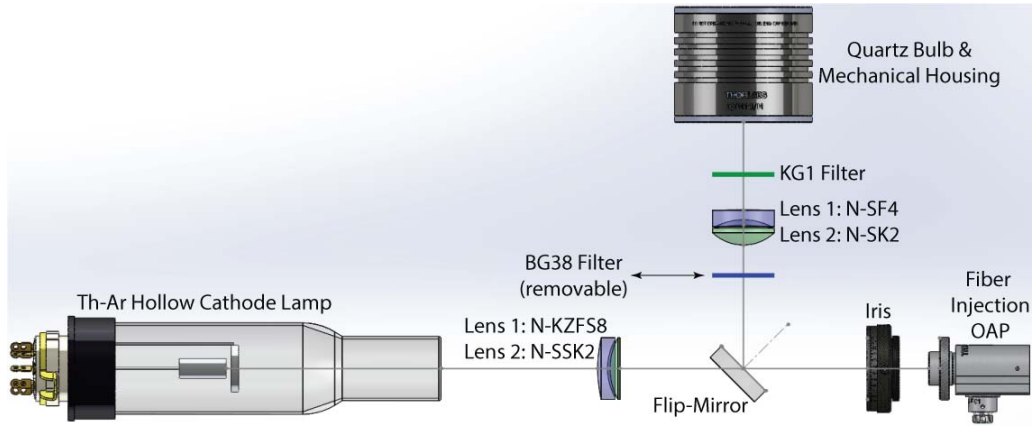

Figure 24 - Optical layout of the calibration unit. Light from either a Th-AR lamp or filtered quartz bulb is focused onto a calibration fiber that feeds the light to the front-end module where it is injected into the science fiber.

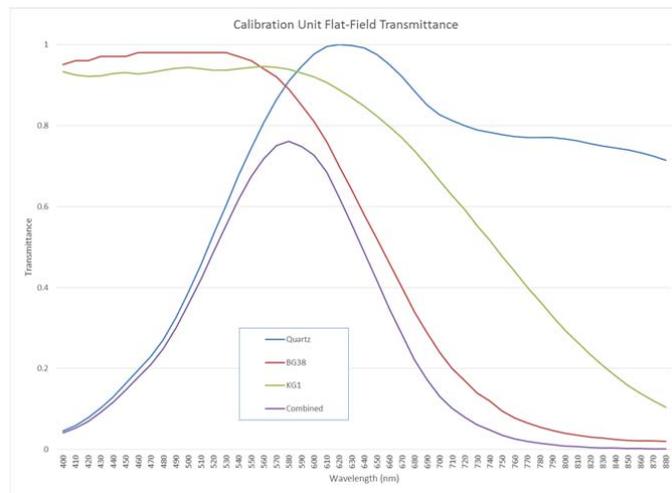

Figure 25 - Transmittance of the calibration unit flat-field source. The quartz bulb light source is filtered through a combination of KG1 and BG38 filters to provide adequate SNR across the entire instrument bandpass in a single exposure.

The filter mechanism was designed so that a combination of flats could be obtained with filtered and unfiltered light from the quartz lamp; the intent was to optimize both red and blue wavelengths of light for flat-fielding. However, during commissioning it was found that data reduction was much more effective using a single quartz flat with a KG1 filter in the system and this filter is currently located in front of the filter optics. Figure 25 shows the transmission curves for the bare quartz bulb and the filtered BG38 and KG1 light.

Light from the calibration unit is fed to the Front-End Module (FEM), located on the telescope Cassegrain instrument port, through a circular fiber. At the FEM, the calibration light is injected into the same octagonal fiber that is used to feed science targets to the spectrograph. Figure 26 shows a cross section of the calibration injection mechanism located in the FEM. A computer controlled flip mirror is deployed to direct the calibration light into the science fiber when calibration images are desired.

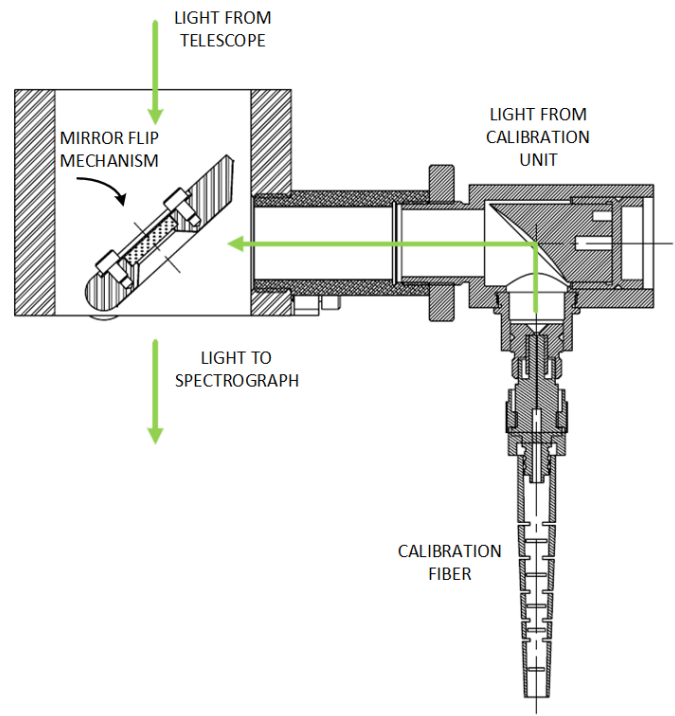

Figure 26 – Calibration injection mechanism located in the FEM on the telescope. Calibration light is fed to the spectrograph through the same fiber as the science objects using a flip mirror mechanism.

## 4.2. *Electronics & Control*

The calibration unit was designed with convenience, reliability, and supportability in mind. It is built on a small optical bench and uses commercially available hardware components and standard software interfaces for control. The calibration unit is enclosed in a standard Thorlabs optical enclosure with sliding doors for service access and black panels to minimize external light contamination. Figure 27 shows the calibration unit with the enclosure sides and top removed.

Lamp power and the filter/flip mirror mechanisms are controlled by Advantech Ethernet I/O modules that communicate using a Modbus TCP protocol. A custom PCB was designed to provide high voltage for the ThAr lamp, and motor control of the calibration injection mechanism flip mirror located in the FEM. The calibration unit is operated at the device level using National Instrument LabVIEW software, which is linked to the user level interfaces for seamless operation and monitoring of the calibration functions. The status of the calibration system is written to image headers for each exposure.

Although most control on the calibration unit is available to the user, there are some functions built into the device level control to protect or preserve the hardware. Limit switches associated with motions, such as the flip mirror insertion in the FEM, are interlocked to the motor drive. There are timeouts on the lamp power enables to prevent premature ageing of the lamps if they are inadvertently left on.

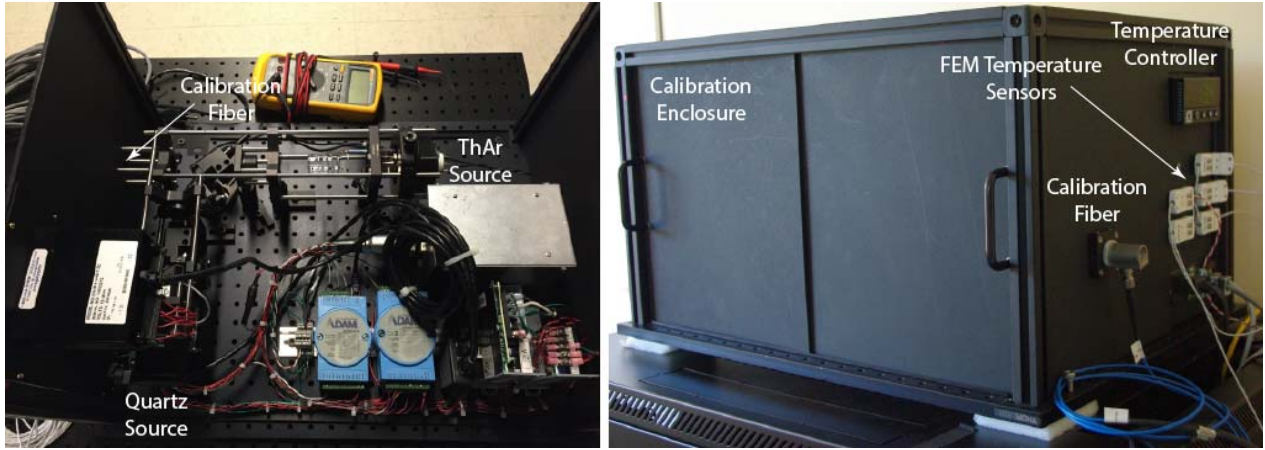

Figure 27 - The MAO spectrograph calibration unit assembly. (left image) Enclosure covers removed showing ThAr lamp in the upper right, the quartz lamp in the lower left, fiber couples outside of enclosure in the upper left, electronics and power supplies are located in the lower right. (right image) Calibration enclosure showing external connections for calibration fiber and temperature sensors, and temperature controller display panel.

## 5. Instrument Control & Data Reduction

### 5.1. *Instrument Control Software*

The instrument control software runs on two computers operating Linux. The software was designed to provide the user with a minimal graphical interface capable of complete spectrograph control. Four separate applications are used: a detector application for taking exposures, a control application for setting up the spectrograph, an environmental application for monitoring temperature and pressure, and an autoguiding application. All instrument control and data acquisition is handled on one computer with communication between applications done through a TCP/IP protocol. A separate computer is dedicated to the autoguider which communicates with the telescope control system via a Modbus TCP protocol. Figure 28 shows a top-level diagram of the instrument control software suite.

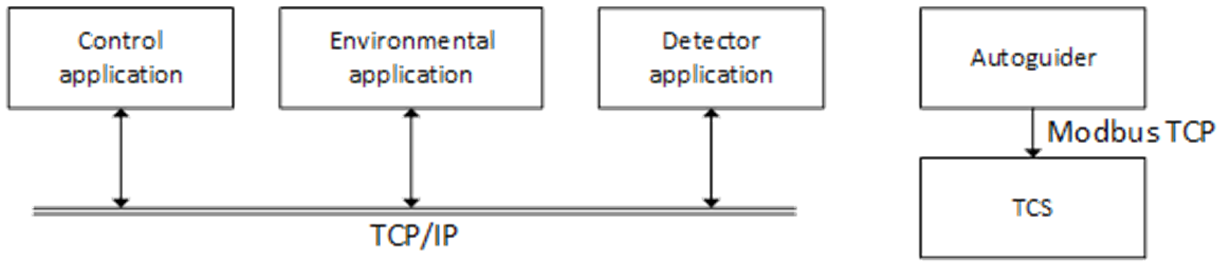

Figure 28 – Top level diagram of instrument control software. Communication between applications is through TCP/IP while the autoguider uses Modbus TCP for communicating with the telescope control system.

#### 5.1.1. *Detector Application*

The detector is controlled and operated with an Astronomical Research Cameras Inc. controller and its associated Owl graphical user interface (GUI). This was chosen as it provided an off-the-shelf solution with a developed interface that offered the ability for scripting on the user end. The GUI provides an interface to the controller and the necessary functionality for making observations. A custom post exposure script was provided for populating the FITS header with the instrument configuration and status. The script queries the control applications using the TCP/IP protocol.

### 5.1.2. *Control and Environmental Application*

A modular approach was taken with the instrument control software; individual modules are associated with a single component of the instrument (calibration lamps, slit mask, temperature). The top-level control applications launch all modules; each module self-initializes once started and reports its status to the main application. This allows for future upgrades to the instrument, e.g. a filter wheel, with minimal effort in adapting the control software. The primary function of each module is presented to the user in a single unified LabVIEW GUI; control of individual modules is offered in an alignment mode. Communication between the GUI and modules is performed through TCP/IP interface to allow for scripting of the instrument though this was not tested. The environmental application monitors the BEM temperature and pressure and controls the FEM temperature. It was separated from the instrument control application for background operation and it logs all data to disk.

## 5.2. **Electronics Architecture**

The spectrograph electronics were designed for performance and reliability. All components are commercially available and, except for the autoguider, make use of either third party or native LabVIEW drivers.

### 5.2.1. *FEM*

Located in the FEM is the calibration injection flip mirror, a heater circuit, a temperature sensor, and the autoguiding camera. The calibration flip mirror is automated such that when calibration frames are to be taken, the flip mirror is deployed without user interaction. When deployed the mirror directs collimated light from the calibration fiber assembly into the science fiber. A Firgelli DC motor extends to insert the mirror and is controlled via a relay module in the calibration box.

A resistive heater circuit comprised of two legs of three $10\Omega$ resistors in series provides approximately 10W of power to keep the FEM above the dew point. The circuit is powered by an Omega temperature controller using an RTD located near the focal plane mirror for temperature sensing. The RTD and heater circuit both have leads extending to the Omega controller mounted in the calibration box.

### 5.2.2. *BEM*

The BEM houses a slit mask stage, the focus mechanism, a pressure transducer, and a set of temperature sensors. The desired slit sets the entrance to spectrograph; it is therefore key that the translational stage positioning the slit mask have high precision. A Physiks Instruments translational stage is tasked with locating the mask. The mask is comprised of a 200μm hole in which the beam is left un-vignetted for the low resolution mode (30k) and slits with widths of 135μm and 100μm for the 45k and 60k resolution modes respectively. There is an additional pinhole in the slit mask for alignment purposes only. All slit mask positions are set by variables in a configuration file and selected by the user from the main GUI.

A Thorlabs, lead screw type DC motor interfaces with the second doublet of the camera barrel assembly and is used to focus the spectrum on the detector. The lead screw pushes against a spring assembly designed such that pressure on a single point results in a concentric movement of the entire lens assembly. It was determined that the focus motor would not be needed for nightly science operations and was not integrated into the main control GUI. A command line interface was provided for scripting purposes by the detector GUI.

The BEM environment is passively stabilized. The spectrograph is housed in a light tight enclosure made of insulation panels designed such that if the spectrograph room is controlled to ±2C, the spectrograph itself is stable at the ±1C level. Temperature sensors were placed throughout the spectrograph in order to monitor its stability. A pressure transducer was placed near the echelle grating to monitor the atmospheric pressure within the spectrograph enclosure. All temperatures and the pressure are logged to disk and written to the corresponding keywords in the FITS headers. Temperature stability was verified via finite element analysis.

### 5.2.3. *Detector Controller*

The detector is an e2v CCD231-84 which has square 15μm pixels arranged in a 4096x4112 format. The detector was AR coated with a uniform broadband astronomy coating and mounted in a cryostat cooled with a closed cycle refrigerator cooling system from Advanced Research Systems. Manufacturer tests reported read noises ranging from three to six electrons depending on the readout speed. The array temperature is set with the Astronomical Research Cameras controller and was delivered with an operating temperature of -104C.

## 5.3. **Alignment Software**

A full suite of algorithms written in IDL were delivered with the spectrograph. These scripts are used in diagnosing and adjusting the alignment of the instrument, useful for intermittently checking the alignment after a major modification to the instrument such as servicing of the cryostat, changing of the science fiber, etc. The focus program will be operated routinely each night to check focus. All code is available on GitHub.

### 5.3.1. *Blaze Alignment*

The blaze peak must be centered on the detector to ensure that the full free spectral range is on the chip. This is performed by taking an unsaturated flat field and tracing a single echelle order. The order is fitted with a polynomial on the detector and summed in the cross-dispersion direction. A Gaussian fit is used to center the one-dimensional order with the peak corresponding to the location of the blaze peak. The cryostat is then translated in the dispersion direction to center the blaze peak on the chip.

To center the detector on the spectrum in the cross-dispersion direction, a monochromator was used to determine the cryostat translation. The monochromator, which has a 6nm tunable bandpass, was set to emit 400nm light and injected into spectrograph. This traces the shortest wavelength of interest on the echelle format, allowing for centering of the spectral format on the chip by translating the cryostat. The monochromator was set to 880nm and it was verified the entire bandpass of interest, 400nm to 880nm, was on the detector. In the absence of a monochromator, an LED with the appropriate wavelengths could have been used.

### 5.3.2. *Detector Rotation*

The rotation of the spectral format was also checked. Because the spectral format wraps in adjacent orders, single lines were identified in two orders of the ThAr spectrum (the lines fall blueward on one order and redward on the next order). The x,y positions of the matching spectral lines were determined by fitting Gaussian profiles and the cryostat was rotated so that the same spectral lines in adjacent orders fell on the same row of the CCD.

### 5.3.3. *Focus and Tilt Adjustment*

Detector focus is implemented by adjusting the second doublet in the camera barrel lens assembly using a DC motor. After a rough "by eye" focus has been achieved, a series of ThAr spectra are taken, systematically stepping through a coarse grid of focus positions. The focus program calculates the FWHM of a user-selected line list and displays this information for both the pixel rows and columns. Once the region of best focus has been determined, a series of ThAr spectra are obtained with a finer grid of focus positions. A polynomial fit to the FWHM measurements as a function of motor position demonstrates the best focus position. The focus motor steps should always proceed in a single direction to avoid backlash in the motor. If any trends in the FWHM are observed along either the columns or rows this tilt in the detector with respect to the focal plane of the optics can be corrected by shimming the cryostat.

## 5.4. **Data Reduction Pipeline**

A full data reduction pipeline was delivered with the instrument that included bias subtraction and made use of quartz calibrations for flat fielding and ThAr spectra for wavelength calibration. The output of the data reduction pipeline is the extracted spectrum in a three dimensional FITS file; the wavelength of each pixel, the flux for each pixel, and the echelle order. All reduction code is hosted on GitHub.

### 5.4.1. Wavelength Calibration

Wavelength calibration of the spectrograph is carried out with a ThAr emission lamp. The first time that this calibration is carried out, the process involves manual identification of the ThAr spectral lines. However, after the initial calibration, the wavelength solution is saved and used as an initial guess (small perturbations in line positions are not a problem) and the wavelength calibration runs as an automated process. The initial calibration process may need to be carried out after any major modifications, such as cryostat servicing, science fiber replacement, re-alignment, etc.

### 5.4.2. Automation

The purpose of the pipeline is to calibrate wavelengths and extract 1-D spectra from a two dimensional echelle format. Once an adequate wavelength mapping is established on file, a single command issued from the IDL reduction environment will calibrate and extract the data. The pipeline reads a nightly log sheet that is specifically formatted to identify files corresponding to bias frames, ThAr, quartz, and science observations. All files must be located in the appropriate folders within an established directory hierarchy. These standard pipeline strategies provide a robust and efficient automation for data analysis.

## 6. Laboratory Results

In this section, we summarize the results of the throughput and preliminary alignment of the spectrograph at Yale University. The BEM was completely assembled and aligned to test both the instrument control software and data reduction pipeline.

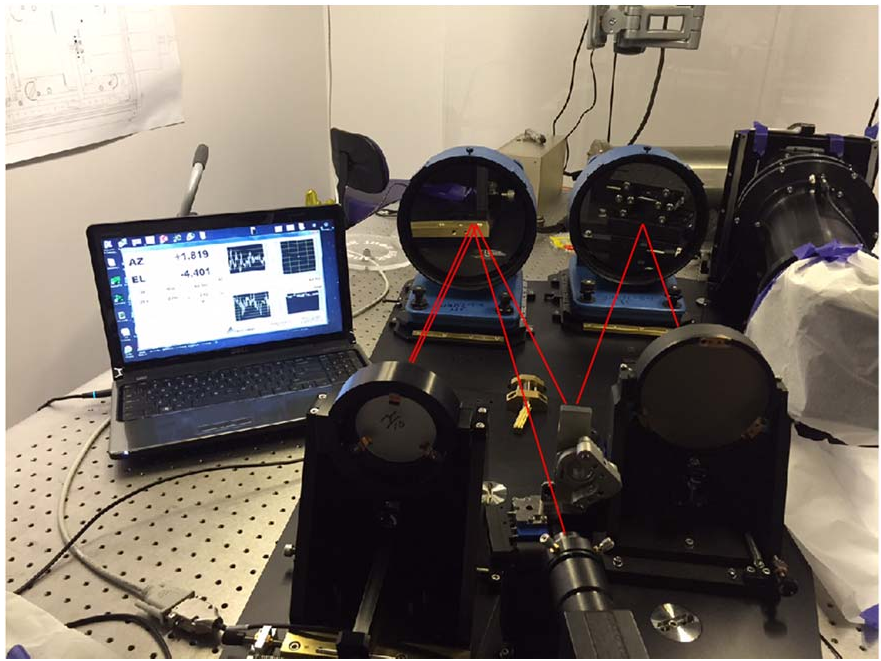

Figure 29 – Spectrograph alignment in retro-reflect mode.

### 6.1. Throughput

Table 4 lists the estimated throughput for the VUES spectrograph at two laser wavelengths, used in the lab to verify performance. For each wavelength in the table actual measured mirror coating reflectance, anti-reflection coating performance, and material bulk transmission values are used. The echelle efficiency was measured by Newport for three orders with blaze peaks near 370, 633, and 898 nm. From these curves, the efficiency at the laser wavelengths was estimated based on its location within the blaze function.

Table 4 – Tabulation of the estimated instrumental throughput for two laser wavelengths used in the lab to verify BEM throughput. Measured values for the BEM were 36.6% and 29.6% for 543 and 652 nm respectively. Given timing constraints for the delivery and installation of the instrument, Yale was not required to provide total throughput measurements.

| Error Source | Throughput | | | | | |
|---|---|---|---|---|---|---|
| | 543 nm | | | 652 nm | | |
| | L1 | L2 | L3 | L1 | L2 | L3 |
| Total VUES Throughput | 0.224 | | | 0.226 | | |
| FEM | | 0.571 | | | 0.668 | |
| Fiber Feed Mirror Rejection | | | 0.780 | | | 0.780 |
| Fiber Camera (4 Elements) | | | 0.817 | | | 0.904 |
| Fiber Coupling at FEM | | | 0.980 | | | 0.980 |
| Octagonal Science Fiber (20 m) | | | 0.914 | | | 0.966 |
| BEM | | 0.393 | | | 0.338 | |
| Fiber Coupling at BEM | | | 0.960 | | | 0.960 |
| Fiber OAPs | | | 0.960 | | | 0.951 |
| Main Collimator (two passes) | | | 0.960 | | | 0.951 |
| Echelle Grating | | | 0.700 | | | 0.600 |
| Transfer Mirror | | | 0.980 | | | 0.975 |
| Transfer Collimator | | | 0.980 | | | 0.975 |
| Fold Mirror | | | 0.980 | | | 0.975 |
| Prism | | | 0.911 | | | 0.960 |
| Camera (5 Elements) | | | 0.777 | | | 0.882 |
| Detector QE | | | 0.951 | | | 0.826 |

Throughput at the two laser wavelengths for the BEM was measured in the lab using a power meter in place of the CCD detector. With a factor applied to the measurements to account for the detector QE, the BEM throughput was measured to be 36.6% and 29.6% for 543 and 652 nm respectively. At these wavelengths the measured throughput was 3-4% below the estimated values which we feel is quite good considering that ideal values were used for the estimate. The FEM throughput was not measured in the lab due to constraints in the delivery and installation schedule of the instrument.

## 6.2. Spectral Resolving Power

The spectral resolving power was determined from the FWHM of the ThAr spectral line spread functions sampled over the full spectral format. Table 5 summarizes the results for the three different slit positions. The low resolution mode has a slit mask width of 500 µm, the medium resolution is 135 µm, and the high is 100 µm.

Table 1 – Lists the measured line FWHM in pixels, and the spectral resolving power for the three slit mask positions.

| Slit Mask Width | Line FWHM (pixels) | Resolving Power ($\lambda/\Delta\lambda$) |
|---|---|---|
| 500 µm (low) | 7.3 | 36,570 |
| 135 µm (medium) | 5.7 | 51,989 |
| 100 µm (high) | 4.1 | 67,408 |

### 6.3. *Alignment*

An autocollimator was used for alignment of the spectrograph reflective optics, up to the cross-disperser and camera barrel assembly. The echelle grating was replaced with a flat mirror and the fold mirror was rotated into a retro-reflect mode to direct light coming from the transfer OAP back on itself (Figure 29). The autocollimator was aligned with a pair of fiducials to the optical path of the main collimating OAP; the echelle flat was then aligned to the autocollimator. This step was repeated for the transfer OAP/fold mirror axis. Both OAPs were then aligned to their respective retro-reflect mirrors. The autocollimator was then injected at f/10 into the spectrograph and the echelle flat rotated out of retro-reflect mode to transfer the focus of the main collimator to the transfer collimator. The transfer mirror could then be aligned resulting in the spectrograph operating in a complete retro-reflect mode, this configuration is shown in Figure 29. The autocollimator reports an alignment error of less than 5 arcseconds in both azimuth and elevation.

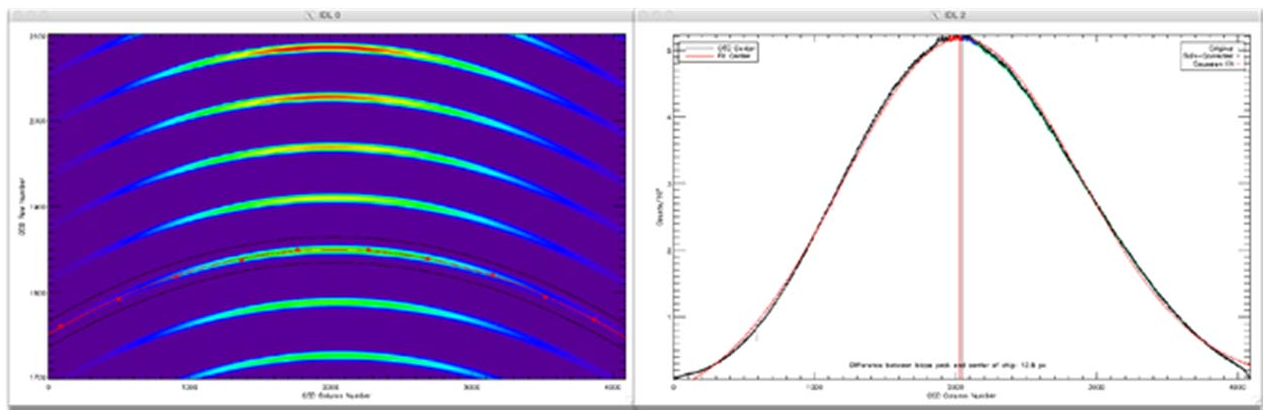

Figure 30 – Centering of the blaze function on the detector. The blaze peak is ~12 pixels from the center.

Alignment of the camera barrel assembly and detector requires the aide of the alignment algorithms discussed in section 5.3. Figures 30 through 32 show the results of the algorithms and what is presented to the user. The fold mirror was rotated to center the beam on the cross-disperser and direct light to the detector. The cryostat was then translated to center the blaze function on the detector, Figure 30; rotated such that repeated wavelengths on adjacent orders were optimally on the same row, Figure 31; and focus/tilt re-checked, Figure 32.

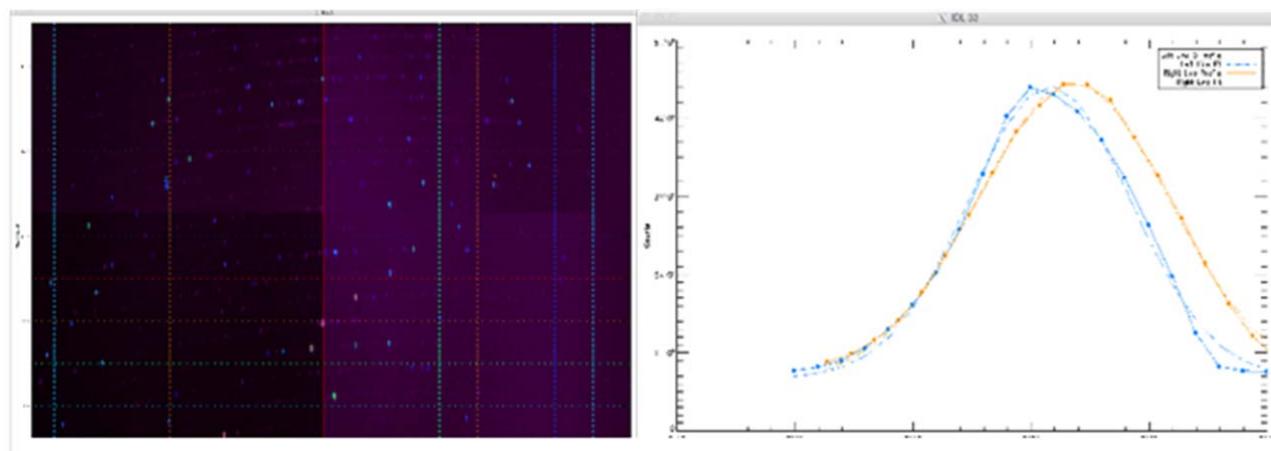

Figure 31 – Cryostat rotation alignment. A spectral line repeats on the adjacent order two rows above the first.

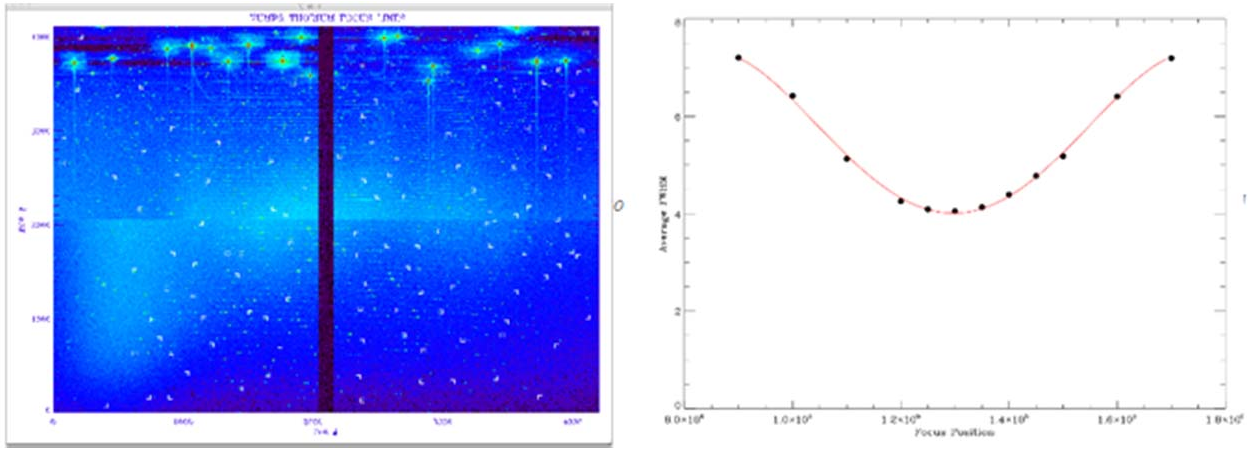

Figure 32 – Detector focus sequence showing position of best focus.

### 6.4. *Solar Spectrum*

A solar spectrum was taken after the spectrograph was completely aligned. After the initial wavelength calibration sequence, the solar spectrum was reduced with the pipeline. Shown in Figure 33 is the reduced solar spectrum with the blaze function fitted and removed.

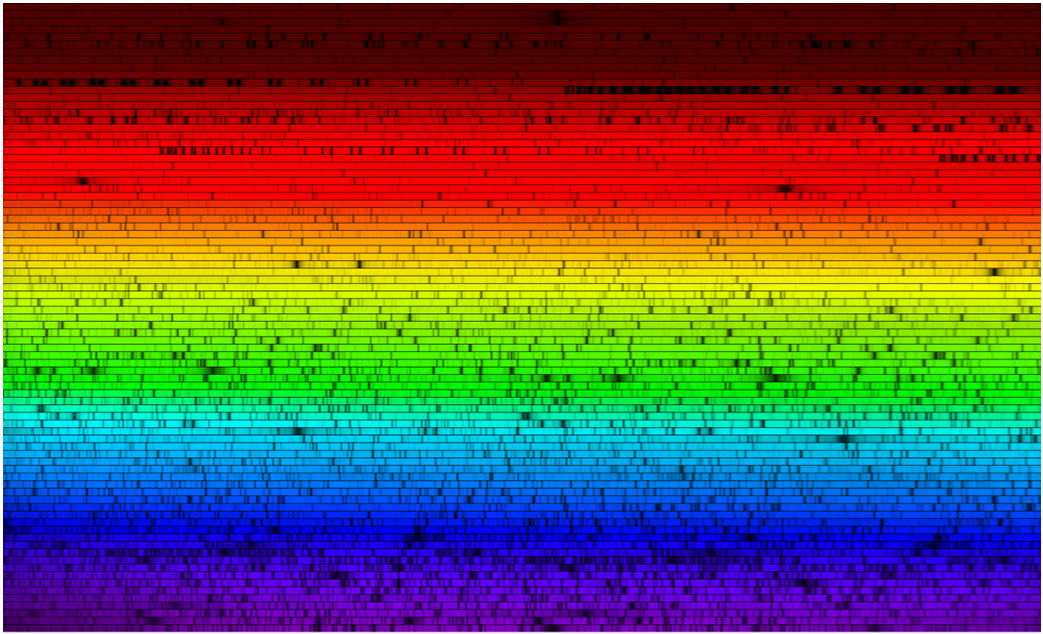

Figure 33 – Extracted and wavelength calibrated solar spectrum taken at Yale University. The blaze function was fit and removed from each extracted order.

## 7. Conclusion

The approach of using a modular template design for the spectrograph optomechanical assembly along with commercially available components proved to be an effective and reliable method for the construction of a spectrograph in a remote location. In addition, the automation of functions and scripted observing allows for ease of operation, and the included reduction pipeline provides consistent and immediate wavelength calibrated spectra.

**Acknowledgments**

Infrastructure and lab support from Yale University is gratefully acknowledged. We also acknowledge support from Bioeksma and Ramunas Diliautas. We also acknowledge Julius Sperauskas and Vilnius University, who were our collaborators in Lithuania. Andrei Tokovinin of the Cerro Tololo Inter-American Observatory is also acknowledged for project advice and participation in design reviews.